\documentclass[10pt,conference]{IEEEtran}
\IEEEoverridecommandlockouts

\usepackage{mdframed}

\usepackage{algorithm}
\usepackage{cite}
\usepackage{amsmath,amssymb,amsfonts}
\usepackage{algorithmic}
\usepackage{graphicx}
\usepackage{textcomp}
\usepackage{xcolor}
\usepackage{hyperref} 
\usepackage{balance} 

\definecolor{darkgreen}{RGB}{0,160,0}

\usepackage{listings}
\usepackage{xcolor}
\usepackage{enumitem}

\usepackage{url}

\usepackage{threeparttable,booktabs}
\usepackage{array}
\usepackage{makecell}
\usepackage{cellspace}
\usepackage{colortbl}
\usepackage{siunitx}
\def\hour{\mathrm{h}}
\def\minute{\mathrm{m}}
\def\second{\mathrm{s}}
\usepackage{subcaption}

\usepackage{booktabs}

\usepackage{pifont}

\newcommand{\xmark}{\ding{55}}
\usepackage{graphicx}
\usepackage{multirow}
\usepackage{textcomp}
\usepackage{tcolorbox}
\usepackage{todonotes}

\usepackage[utf8]{inputenc}
\definecolor{eclipseStrings}{RGB}{42,0,255}
\definecolor{eclipseKeywords}{RGB}{127,0,85}
\definecolor{bgcolor}{HTML}{f5f4f2}
\colorlet{numb}{magenta!60!black}
\usepackage{tcolorbox}
\usepackage{float}
\usepackage{cleveref}

\setlist[enumerate,1]{left=0pt, itemsep=0pt, parsep=0pt, topsep=0pt}

\usepackage{listings, xcolor}

\definecolor{verylightgray}{rgb}{.97,.97,.97}

\lstdefinelanguage{Solidity}{
	keywords=[1]{anonymous, assembly, assert, balance, break, call, callcode, case, catch, class, constant, continue, constructor, contract, debugger, default, delegatecall, delete, do, else, emit, event, experimental, export, external, false, finally, for, function, gas, if, implements, import, in, indexed, instanceof, interface, internal, is, length, library, log0, log1, log2, log3, log4, memory, modifier, new, payable, pragma, private, protected, public, pure, push, require, return, returns, revert, selfdestruct, send, solidity, storage, struct, suicide, super, switch, then, this, throw, transfer, true, try, typeof, using, value, view, while, with, addmod, ecrecover, keccak256, mulmod, ripemd160, sha256, sha3}, 
	keywordstyle=[1]\color{blue}\bfseries,
	keywords=[2]{address, bool, byte, bytes, bytes1, bytes2, bytes3, bytes4, bytes5, bytes6, bytes7, bytes8, bytes9, bytes10, bytes11, bytes12, bytes13, bytes14, bytes15, bytes16, bytes17, bytes18, bytes19, bytes20, bytes21, bytes22, bytes23, bytes24, bytes25, bytes26, bytes27, bytes28, bytes29, bytes30, bytes31, bytes32, enum, int, int8, int16, int24, int32, int40, int48, int56, int64, int72, int80, int88, int96, int104, int112, int120, int128, int136, int144, int152, int160, int168, int176, int184, int192, int200, int208, int216, int224, int232, int240, int248, int256, mapping, string, uint, uint8, uint16, uint24, uint32, uint40, uint48, uint56, uint64, uint72, uint80, uint88, uint96, uint104, uint112, uint120, uint128, uint136, uint144, uint152, uint160, uint168, uint176, uint184, uint192, uint200, uint208, uint216, uint224, uint232, uint240, uint248, uint256, var, void, ether, finney, szabo, wei, days, hours, minutes, seconds, weeks, years},	
	keywordstyle=[2]\color{teal}\bfseries,
	keywords=[3]{block, blockhash, coinbase, difficulty, gaslimit, number, timestamp, msg, data, gas, sender, sig, value, now, tx, gasprice, origin},	
	keywordstyle=[3]\color{violet}\bfseries,
	identifierstyle=\color{black},
	sensitive=false,
	comment=[l]{//},
	morecomment=[s]{/*}{*/},
	commentstyle=\color{gray}\ttfamily,
	stringstyle=\color{red}\ttfamily,
	morestring=[b]',
	morestring=[b]"
}

\lstdefinelanguage{json}{
    basicstyle=\footnotesize, 
    commentstyle=\color{gray}\ttfamily, 
    stringstyle=\color{blue}\ttfamily, 
    keywordstyle=\color{blue}\bfseries, 
    numbers=left,
    numberstyle=\footnotesize, 
    stepnumber=1,
    numbersep=3pt,
    showstringspaces=false,
    breaklines=true,
    backgroundcolor=\color{bgcolor},
    string=[s]{"}{"},
    morestring=[b]',
    morestring=[b]",
    comment=[l]{:\ "}, 
    morecomment=[l]{:"},
    literate=
        *{0}{{{\color{numb}0}}}{1}
         {1}{{{\color{numb}1}}}{1}
         {2}{{{\color{numb}2}}}{1}
         {3}{{{\color{numb}3}}}{1}
         {4}{{{\color{numb}4}}}{1}
         {5}{{{\color{numb}5}}}{1}
         {6}{{{\color{numb}6}}}{1}
         {7}{{{\color{numb}7}}}{1}
         {8}{{{\color{numb}8}}}{1}
         {9}{{{\color{numb}9}}}{1}
         {:}{{{\color{red}:}}}{1} 
         {,}{{{\color{gray},}}}{1} 
}

\lstdefinelanguage{Solidity2}{
  keywords={contract, mapping, function, public, payable, require, bool, uint256, address},
  keywordstyle=\color{blue}\bfseries,
  ndkeywords={pragma, solidity},
  ndkeywordstyle=\color{red}\bfseries,
  identifierstyle=\color{black},
  sensitive=false,
  comment=[l]{//},
  morecomment=[s]{/*}{*/},
  commentstyle=\color{gray}\ttfamily,
  stringstyle=\color{orange}\ttfamily,
  morestring=[b]',
  morestring=[b]"
}

\lstdefinelanguage{Solidity}{
  keywords={contract, int, public, private, function, returns, view},
  keywordstyle=\color{blue}\bfseries,
  ndkeywords={},
  sensitive=false,
  comment=[l]{//},
  morecomment=[s]{/*}{*/},
  commentstyle=\color{gray}\ttfamily,
  stringstyle=\color{orange}\ttfamily,
  morestring=[b]',
  morestring=[b]",
  numbersep=3pt,
}

\tcbuselibrary{listingsutf8}

\newtcblisting{diffbox}[1][]{
  colback=gray!10,  
  colframe=black!0,
  listing only,
  listing options={language=Solidity},
  boxsep=0pt,        
  left=0pt, right=0pt, top=0pt, bottom=0pt,  
  beforeafter skip=0pt,  
  #1
}

\newcommand{\toolname}{\textsc{SoliDiffy~}}
\newcommand{\othertool}{{Difftastic~}}

\newcommand{\rqOne}{How does the performance of \toolname compare to the most-closely related tool, Diffstastic?}
\newcommand{\rqTwo}{How does \toolname perform when there are multiple changes in the smart contract source code?}
\newcommand{\rqThree}{How does the type of syntactic changes in file affect the performance of \textsc{SoliDiffy}?}
\newcommand{\rqFour}{How does \toolname perform against line differencing on commit history of real-world smart contracts?}

\makeatletter
\newcommand{\TODO}[1]{%
  \bgroup
  \def\@tempa{#1}%
  \expandafter\textcolor\expandafter{red}{\@tempa}%
  \GenericWarning{}{LaTeX Warning: TODO: \@tempa}%
  \egroup
}
\makeatother

\makeatletter
\newcommand{\NOTE}[1]{%
  \bgroup
  \def\@tempa{#1}%
  \expandafter\textcolor\expandafter{blue}{\@tempa}%
  \GenericWarning{}{LaTeX Warning: NOTE: \@tempa}%
  \egroup
}
\makeatother

\def\BibTeX{{\rm B\kern-.05em{\sc i\kern-.025em b}\kern-.08em
    T\kern-.1667em\lower.7ex\hbox{E}\kern-.125emX}}
\begin{document}

\title{\textsc{SoliDiffy}: AST Differencing for Solidity Smart Contracts}



\author{
    \IEEEauthorblockN{
        Mojtaba Eshghie\IEEEauthorrefmark{1}, 
        Viktor Åryd\IEEEauthorrefmark{1}, 
        Cyrille Artho\IEEEauthorrefmark{1}, and        
        Martin Monperrus\IEEEauthorrefmark{1}, 
    } 
    \IEEEauthorblockA{
        \IEEEauthorrefmark{1}\textit{KTH Royal Institute of Technology}, Stockholm, Sweden \\
        Email: \{eshghie, viktoraaryd, monperrus, artho\}@kth.se
    }
}

\maketitle

\begin{abstract}
Structured code differencing is the act of comparing the hierarchical structure of code via its abstract syntax tree (AST) to capture modifications. 
AST-based source code differencing enables tasks such as vulnerability detection and automated repair where traditional line-based differencing falls short. We introduce \textsc{SoliDiffy}, the first AST differencing tool for Solidity smart contracts with the ability to generate an edit script that soundly shows the structural differences between two smart-contracts using \emph{insert}, \emph{delete}, \emph{update}, \emph{move} operations. 
In our evaluation on \num{353262} contract pairs, \toolname\ achieved a $96.1\%$ diffing success rate, surpassing the state-of-the-art, and produced significantly shorter edit scripts. Additional experiments on $925$ real-world commits further confirmed its superiority compared to Git line-based differencing. \toolname provides accurate representations of smart contract evolution even in the existence of multiple complex modifications to the source code.
\toolname is publicly available at~\href{https://github.com/mojtaba-eshghie/SoliDiffy}{https://github.com/mojtaba-eshghie/SoliDiffy}.
\end{abstract}


\definecolor{responseColor}{HTML}{3888D6}
\definecolor{conditioncolor}{HTML}{CC7F0C}
\definecolor{milestoneColor}{HTML}{BA1FE5}
\definecolor{includeColor}{HTML}{2FA71F}
\definecolor{excludeColor}{HTML}{C10300}
\definecolor{noresponseColor}{HTML}{8c6026} 
\definecolor{valueColor}{HTML}{8c8c8c} 

\newcommand{\evalexp}[2]{\ensuremath{[[#1]]_{#2}}}
\newcommand{\DCR}{DCR\xspace}
\newcommand{\DCRR}{DCR$^*$\xspace}
\newcommand{\DCRL}{DCR$^\nu$\xspace}
\newcommand{\DCRB}{DCR$^!$\xspace}
\newcommand{\effectoff}[2]{\ensuremath{#1\cdot #2}}
\newcommand{\inputact}[2]{\ensuremath{?(#1, #2)}}
\newcommand{\outputact}[3]{\ensuremath{!(#1, #2, #3)}}
\newcommand{\outact}[2]{\ensuremath{!(#1, #2)}}
\newcommand{\intact}[2]{\ensuremath{*(#1, #2)}}
\newcommand{\ioactions}{\ensuremath{\mathsf{IO_{A,D}}}}
\newcommand{\inoutactions}{\ensuremath{\mathsf{IO_{A}}}}
\newcommand{\interact}[3]{\ensuremath{(#1, #2 \rightarrow #3)}}
\newcommand{\actt}[3]{\ensuremath{#1(#2,#3)}}
\newcommand{\interactdata}[4]{\ensuremath{(#1, #2 \stackrel{#3}{\longrightarrow} #4)}}
\newcommand{\rref}{\sqsubseteq}
\renewcommand\t{\ensuremath{\mathsf{t}}}
\newcommand\f{\ensuremath{\mathsf{f}}}
\newcommand{\VAL}{\ensuremath{V}} 
\newcommand{\natinf}{\ensuremath{\infty}}

\newcommand{\lab}{\ensuremath{l}}
\newcommand{\valuerel}{{\color{valueColor} \ensuremath{\mathrel{\rightarrow\!\!=}}}}
\newcommand{\conditionrel}{{\color{conditioncolor} \ensuremath{\mathrel{\rightarrow\!\!\bullet}}}}
\newcommand{\responserel}{{\color{responseColor}\ensuremath{\mathrel{\bullet\!\!\rightarrow}}}}
\newcommand{\milestonerel}{{\color{milestoneColor} \ensuremath{\mathrel{\rightarrow\!\!\diamond}}}}
\newcommand{\includerel}{{\color{includeColor} \ensuremath{\mathrel{\rightarrow\!\!\textsf{+}}}}}
\newcommand{\excluderel}{{\color{excludeColor} \ensuremath{\mathrel{\rightarrow\!\!\textsf{\%}}}}}

\newcommand{\noresponserel}{{\color{noresponseColor}\ensuremath{\mathrel{\bullet\!\!\!\rightarrow\!\!\!\times}}}}

\newcommand{\gconditionrel}[2]{{\color{conditioncolor}\ensuremath{\stackrel{[#1]}{\mathrel{\rightarrow\!\!\bullet}}_{d}}}}
\newcommand{\gresponserel}[2]{{\color{responseColor}\ensuremath{\stackrel{[#1]}{\mathrel{\bullet\!\!\rightarrow}}_{#2}}}}
\newcommand{\gmilestonerel}[1]{{\color{milestoneColor}\ensuremath{\stackrel{[#1]}{\mathrel{\rightarrow\!\!\diamond}}}}}
\newcommand{\gincluderel}[1]{{\color{includeColor}\ensuremath{\stackrel{[#1]}{\mathrel{\rightarrow\!\!+}}}}}
\newcommand{\gexcluderel}[1]{{\color{excludeColor}\ensuremath{\stackrel{[#1]}{\mathrel{\rightarrow\!\!\%}}}}}
\newcommand{\gnoresponserel}[1]{{\color{noresponseColor}\ensuremath{\stackrel{[#1]}{\mathrel{\bullet\!\!\!\rightarrow\!\!\!\times }}}}}

\newcommand{\responses}{\ensuremath{\mathsf{Re}}}
\newcommand{\executed}{\ensuremath{\mathsf{Ex}}}
\newcommand{\included}{\ensuremath{\mathsf{In}}}

\newcommand{\markingset}{\ensuremath{\mathcal{M}}}
\newcommand{\graphs}{\ensuremath{\mathcal{G}}}
\newcommand{\pgraphs}{\ensuremath{{\cal{P}}}}

\newcommand{\genrel}{\ensuremath{\mathrel{\rightarrow}}}
\newcommand{\genrelto}[1]{\genrel\!#1}
\newcommand{\genrelfrom}[1]{#1\!\genrel}
\newcommand{\enable}[2]{\ensuremath{\mathsf{enabled}(#1,#2)}} 
\newcommand{\power}[1]{\ensuremath{{\cal P}(#1)}}
\newcommand{\maxc}{\ensuremath{maxc_G}}
\newcommand{\minr}{\ensuremath{minr_G}}
\newcommand{\miner}{\ensuremath{minRe_G}}
\def\L{\mathsf{L}}

\def\lEx{\mathsf{L_{Ex}}}
\def\lRe{\mathsf{L_{Re}}}
\def\lIn{\mathsf{L_{In}}}
\def\dom{\mathsf{dom}}

\newcommand{\spg}{\ensuremath{sp}} 
\newcommand{\ESG}{\ensuremath{E}}

\newcommand\Ex{\mathsf{Ex}}
\renewcommand\Re{\mathsf{Re}}
\newcommand\In{\mathsf{In}}

\newcommand\Va{\mathsf{Va}}

\newcommand{\commitevent}{\ensuremath{\mathsf{commit}}}
\newcommand{\revealevent}{\ensuremath{\mathsf{reveal}}}
\newcommand{\transactionevent}{\ensuremath{\mathsf{revealtransaction}}}
\newcommand{\placeInEscrowEvent}{\ensuremath{\mathsf{placeInEscrow}}}
\newcommand{\releaseBySenderEvent}{\ensuremath{\mathsf{releaseBySender}}}
\newcommand{\releaseByReceiverEvent}{\ensuremath{\mathsf{releaseByReceiver}}}
\newcommand{\withdrawFromEscrowEvent}{\ensuremath{\mathsf{withdrawFromEscrow}}}

\newcommand{\draftcomment}[3]{{\color{#1}[#3] #2} 
  \PackageWarning{WARNING: Draft comments visible}{#2: #3}}
\newcommand{\gs}[1]{\draftcomment{red}{GS}{#1} }
\newcommand{\wa}[1]{\draftcomment{blue}{WA}{#1} }

\section{Introduction}\label{sec:intro}
Smart contracts are self-executing programs that implement real-world agreements by encoding contract terms directly into code~\cite{wood2014ethereum,szabo2017smart}. These programs operate on blockchain platforms such as Ethereum~\cite{EthereumHome}, enabling trustless transactions without intermediaries. Solidity, a statically-typed language, is the leading choice for smart contract development~\cite{soliditysite}.

Despite the rapid adoption of smart contracts, software engineering tooling for Solidity is still scarce. 
In particular, the Solidity developer community lacks a robust tool for accurately tracking and analyzing code changes, essential to  tasks such as automated smart-contract vulnerability detection~\cite{feist2019slither,manticore,dynamit,xplogen,highguard} and smart-contract  repair~\cite{SGUARD,reenrepair,SCRepair,Aroc,ContractTinker,bobadilla2025}.

While structural differencing tools exist for other programming languages~\cite{falleri_fine-grained_2014}, a good solution for smart contracts does not exist.  Traditional line-based differencing tools, such as Git diff, fail to capture the hierarchical structure of smart contracts, leading to inaccurate change representations. 
Developers working with Solidity can benefit from fine-grained structural differencing in several scenarios. For instance, when upgrading a smart contract, AST-based differencing enables the identification of subtle structural changes that line-based methods might overlook (Section~\ref{sec:motivating-example}). Similarly, automated program repair (APR) \cite{SGUARD,reenrepair,SCRepair,Aroc,ContractTinker} relies on AST differencing to validate patches, document the transformations, and ensure semantic consistency. AST differencing also plays a crucial role in detecting code clones\cite{gao2020checking,chen2021understanding,gao2019smartembed}, where semantically similar yet syntactically different code blocks must be identified.

To address these gaps, we introduce \textsc{SoliDiffy}, a novel AST differencing tool tailored for Solidity smart contracts. Unlike existing line-based differencing approaches, \toolname accurately captures structural modifications using Solidity-specific AST analysis. 
In \textsc{SoliDiffy}, we devise AST pruning and transformation rules to address Solidity constructs and syntax. 

Next, we conduct an extensive experimental study on \num{354187} contract pairs to answer the following research questions:

\begin{itemize}[left=0pt]
\item \textbf{RQ1:} How does \toolname compare to the state of the art? 
In comparison to Difftastic~\cite{hughes_difftastic_2024}, which generates text-based edit scripts, \toolname provides shorter edit scripts and successfully analyzes more contract pairs.
\item \textbf{RQ2:} How does \toolname perform when there are multiple changes in the smart contract source code? Our analysis demonstrates that \toolname remains effective and consistently generates lower edit distances regardless of the number of stacked modifications.
\item \textbf{RQ3:} How does the type of syntactic changes affect the performance of \textsc{SoliDiffy}? We find that \toolname excels in handling complex structural changes, particularly large code block modifications.
\item \textbf{RQ4:} How does \toolname perform against line differencing on real-world commit histories? Our study on Uniswap v4~\cite{UniswapV4core2024} contracts shows that \toolname provides more accurate and structured representations of code evolution compared to Git line-based differencing.
\end{itemize}

This rest if the paper is structured as follows: 
Section~\ref{sec:motivating-example} presents a motivating example.
Section~\ref{sec:background} provides the necessary background. Section~\ref{sec:extending-gt} demonstrates the architecture of \textsc{SoliDiffy}. Section~\ref{sec:experimental-protocols} outlines our experimental protocol, followed by results in Section~\ref{sec:results} and a discussion on key findings and threats to validity in Section~\ref{sec:discussion}. Section~\ref{sec:related-works} reviews related work. Finally, Section~\ref{sec:conclusion} concludes the paper.

\begin{figure}[t]
\centering
\begin{diffbox}
(*@\colorbox{red!20}{-contract SimpleStorage \{}@*)
(*@\colorbox{red!20}{-    uint256 public num;}@*)
(*@\colorbox{green!20}{+contract SimpleStorage\{}@*)
(*@\colorbox{red!20}{-    function set(uint256 \_num) public \{}@*)
(*@\colorbox{red!20}{-        num = \_num;}@*)
(*@\colorbox{red!20}{-    \}}@*)
(*@\colorbox{green!20}{+    uint256    private counter;}@*)
(*@\colorbox{green!20}{+}@*)
(*@\colorbox{green!20}{+    function set(uint256 \_num) public}@*)
(*@\colorbox{green!20}{+    \{}@*)
(*@\colorbox{green!20}{+        counter = \_num;}@*)
(*@\colorbox{green!20}{+     \}}@*)    
(*@\colorbox{red!20}{-    function increment() public \{}@*)
(*@\colorbox{red!20}{-        num += 1;}@*)
(*@\colorbox{red!20}{-    \}}@*)
(*@\colorbox{green!20}{+    function increment() public\{}@*)
(*@\colorbox{green!20}{+        counter += 1;}@*)
(*@\colorbox{green!20}{+     \}}@*)
(*@\colorbox{red!20}{-    function get() public view returns (uint256) \{}@*)
(*@\colorbox{red!20}{-        return num;}@*)
(*@\colorbox{red!20}{-    \}}@*)
(*@\colorbox{green!20}{+    function get( ) public view returns(uint256)\{}@*)
(*@\colorbox{green!20}{+        return counter;}@*)
(*@\colorbox{green!20}{+     \}}@*)
(*@\colorbox{green!20}{+    function reset() public\{}@*)
(*@\colorbox{green!20}{+       counter = 0;}@*)
(*@\colorbox{green!20}{+     \}}@*)
(*@\colorbox{green!20}{+}@*)
}
\end{diffbox}
\caption{Standard line diff of original and modified \emph{SimpleStorage} contract, with added and removed lines highlighted.}
\label{fig:comparison_example}
\end{figure}

\begin{figure*}[hbt!]
    \centering
    \makebox[\textwidth][c]{
    \includegraphics[width=.9\textwidth]{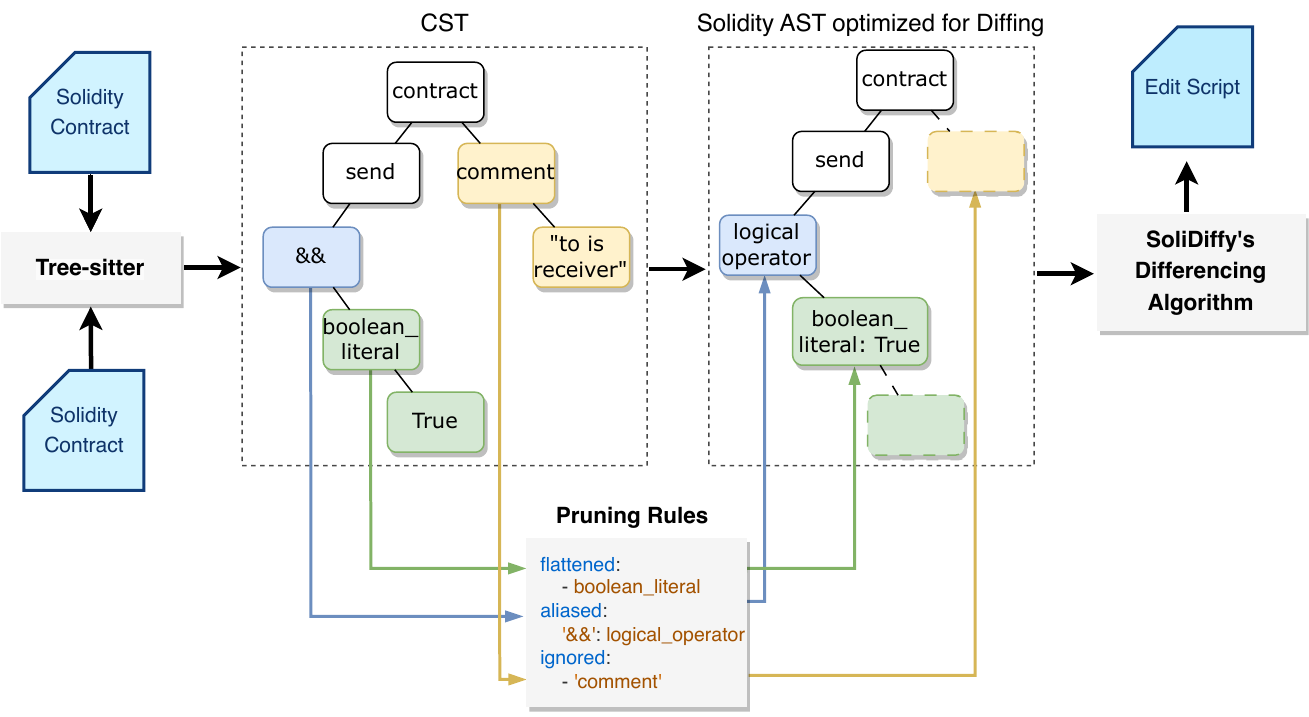}}
    \caption{The design of the \toolname smart contract differencing tool.}
    \label{fig:solidiffy-design}
\end{figure*}

\section{Motivating Example}\label{sec:motivating-example}

Tasks such as program repair and vulnerability detection depend on precisely understanding code modifications. Consider a scenario where a developer modifies the source code of a Solidity smart contract during an upgrade. A basic line-based differencing tool, such as Git diff, will highlight textual changes but lacks the ability to distinguish between superficial edits (e.g., renaming variables) and potentially critical changes affecting security or functionality. Such distinction becomes crucial, especially for automated tools designed to verify or reject suspicious code updates to deployed smart contracts~\cite{LargeScaleProxyPattern,captureDCR}.

As an example, consider the \emph{SimpleStorage} smart contract in Figure~\ref{fig:comparison_example}. A developer renamed the storage variable \lstinline{num} to \lstinline{counter}, modified its visibility from \lstinline{public} to \lstinline{private}, and slightly modified the formatting of existing functions. Standard line-based differencing highlights these as multiple unrelated additions and deletions, as illustrated by numerous confusing red and green lines in the figure. Such textual representation obscures the underlying structural relationship between these changes, making it challenging for reviewers or automated analysis tools to quickly grasp the meaning of the modifications.

In contrast, \toolname produces an edit script (Figure~\ref{fig:edit_script_example}) by precisely pinpointing the semantic modifications made in the source code. 
Unlike Git's line-based diff, \toolname distinguishes between changes that are semantically significant (such as variable visibility) and those that are cosmetic or formatting-related, allowing developers to more quickly understand the real intent behind source code updates.

\section{Background}\label{sec:background}

This section provides the concepts essential for AST-based code differencing.

\subsection{Syntax Trees: Abstract vs. Concrete}
Syntax trees, comprising abstract and concrete syntax trees (ASTs and CSTs), represent the hierarchical structure of source code. ASTs focus on the logical structure by abstracting away syntactic details, making them ideal for tasks like code analysis and transformation~\cite{wile1997abstract}. In contrast, CSTs retain all syntactic elements, including punctuation and keywords, capturing the exact format of the source code, which is essential for precise replication tasks like formatting and refactoring.

\subsection{Code Differencing}\label{sec:code-differencing}
Code differencing is the process of identifying differences between two versions of a codebase. This is crucial for version control, collaborative development, and maintaining code quality. Traditional line-based differencing tools, such as those used in Git~\cite{nugroho2020different}, compare code on a line-by-line basis, which can miss or misinterpret finer structural changes in the code. 

\subsubsection{AST Differencing}\label{sec:ast-diff-background}
AST differencing enhances code comparison by utilizing the hierarchical structure of ASTs. Unlike line-based differencing, AST differencing can identify specific modifications within the code's logical structure. For example, a small change within a line of code can be pinpointed precisely, rather than being treated as a completely new or altered line~\cite{falleri_fine-grained_2014}.

\subsubsection{Edit Scripts}\label{sec:edit-scripts}
A common approach in differencing is to generate an edit script as a sequence of operations required to transform one source code into another (see Figure~\ref{fig:edit_script_example}). The process for AST-based edit scripts typically involves two phases: generating mappings between unchanged nodes of the two ASTs and then deriving an edit script from these mappings. These edit actions reflect modifications to source code. Although generating an optimal edit script is an NP-hard problem~\cite{Chawathe_change_detection}, this method provides a structured way to represent differences between code versions.

\section{\toolname: AST Differencing for Solidity}\label{sec:extending-gt}

This section outlines the core components of \toolname, a novel approach for fine-grained and precise AST differencing of Solidity smart contracts.

Figure~\ref{fig:solidiffy-design} shows \textsc{SoliDiffy}'s architecture. \toolname starts by receiving a pair of Solidity smart contract source codes and generates optimized ASTs for the differencing task (Section~\ref{sec:ast-gen-pruning}). The differencing subsystem then uses the mapping between the ASTs to perform differencing (Section~\ref{sec:mapping-gen}).

\subsection{Pruning Rules}\label{sec:ast-gen-pruning}
CSTs contain a wide range of unnecessary information that may pollute edit scripts. To create ASTs optimized for differencing, \toolname\ employs a series of pruning and transformations on the initial CSTs. This involves flattening nodes, aliasing for consistency, and pruning unnecessary elements, as follows:

\begin{itemize}[left=10pt]
    \item \textbf{Flattening}: Combines child nodes with their parent as a single node. In other words, we stop at one node in the AST and putting as value all the source string corresponding to this node and its children.
    As an example, the constant literal values with type and value are concatenated (first rule in under mapping rules and green sub-trees in Figure~\ref{fig:solidiffy-design}).
    \item \textbf{Aliasing}: Renames node types to facilitate a unified differencing process (second rule ``aliased'' in Figure~\ref{fig:solidiffy-design}).
    \item \textbf{Ignoring}: Removes extraneous nodes, such as formatting elements that do not impact the logical structure of the code (third rule in Figure~\ref{fig:solidiffy-design}).
\end{itemize}


We use specialized configuration of these rules to adapt to Solidity from Gumtree's tree-sitter backend~\cite{falleri_fine-grained_2014,TreesitterIntroduction}.

\subsection{Differencing Algorithm}\label{sec:mapping-gen}
The differencing algorithm of \toolname\ generates mappings between nodes of the two ASTs to identify unchanged and modified elements. Utilizing the efficient algorithm of Gumtree~\cite{falleri_fine-grained_2014}, which has undergone extensive evaluation, \toolname aligns nodes between two given ASTs. This process involves a two-phase mapping strategy:

\begin{itemize}[left=2pt]
    \item \textbf{Top-Down Mapping}: Identifies large, unmodified subtrees to serve as anchors, reducing the complexity of subsequent differencing.
    \item \textbf{Bottom-Up Mapping}: Refines the initial mappings by comparing smaller subtrees and individual nodes, ensuring that all modifications are accurately captured.
\end{itemize}

These mappings are then used to derive an edit script (Section~\ref{sec:SoliDiffy-edit-scripts}).

\begin{figure}[hbt!]
    \centering
    \begin{lstlisting}[language=xml]
<update-node tree="visibility: public [37,43]" label="private"/>
<update-node tree="identifier: num [44,47]" label="counter"/>
<update-node tree="identifier: num [242,245]" label="counter"/>
<update-node tree="identifier: num [98,101]" label="counter"/>
<update-node tree="identifier: num [159,162]" label="counter"/>
<update-node tree="identifier: num [292,295]" label="counter"/>
    \end{lstlisting}
    \caption{Edit script generated by \toolname as a result of diffing task of Figure~\ref{fig:comparison_example}.}
    \label{fig:edit_script_example}
\end{figure}

\begin{figure*}[!t]
    \centering
     \makebox[\textwidth][c]{
     \includegraphics[width=0.99\textwidth]{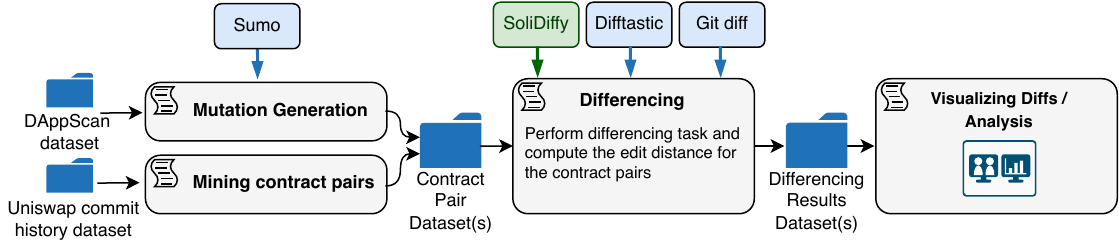}}
    \caption{Pipeline for large-scale generation of contract pair for differencing and subsequent analysis.}
    \label{fig:experiments}
\end{figure*}

\subsection{Edit Script}\label{sec:SoliDiffy-edit-scripts}
\textsc{SoliDiffy}'s edit scripts include four standard operations: \emph{insert}, \emph{delete}, \emph{update}, and \emph{move}. The differencing algorithm prioritizes producing edit scripts that are concise yet fully descriptive of the changes made. 

Fig.~\ref{fig:edit_script_example} shows the edit script generated by \toolname for the differencing task of Fig.~\ref{fig:comparison_example}. The update action in line 1 in this edit script is the edit action to update the visibility of the state variable from \lstinline{public} to \lstinline{private}.  The rest of the update actions propagate the renaming of \lstinline{num} variable to \lstinline{counter}.



\subsection{Implementation}

\toolname{}uses Tree-sitter~\cite{TreesitterIntroduction} to parse Solidity source code into its syntax tree representation. Tree-sitter is an open-source parsing tool that generates CSTs for a wide range of programming languages using a modular grammar framework. We integrated the most recent version of Solidity grammar~\cite{TreesitterIntroduction}.\footnote{\url{https://github.com/JoranHonig/tree-sitter-solidity/blob/a8ed2f5d600fed77f8ed3084d1479998c649bca1/grammar.js}}


\section{Experimental Protocol}\label{sec:experimental-protocols}

We outline the experimental setup to answer our research questions 1--4 in Section~\ref{sec:rqs}. Figure~\ref{fig:experiments} demonstrates this experimental setup, and Section~\ref{sec:protocol} elaborates on the protocol to answer the research questions.  

The experiment is structured as a pipeline that begins with the selection and preparation of our seed datasets of Solidity smart contracts. The seed datasets include 1) DAppSCAN, a large dataset of smart contracts~\cite{zheng_dappscan_2023} and seed dataset from commit history of the Uniswap smart contracts~\cite{UniswapV4core2024} (Section~\ref{sec:datasets}). In the next phase, the contract pairs are generated from our seed datasets. Finally, we run three tools in pairs \toolname, \othertool, and Git line differencing tools according to our protocol for each research question on the generated contract pairs. The upcoming sections provide details of each stage of the experiment, including the dataset preparation, and the methodology for generating and processing source diff pairs.

\subsection{Research Questions}\label{sec:rqs}

The remaining sections of the paper address the following research questions:

\begin{itemize}[left=2pt]

\item \textbf{RQ1:} How does the performance of \toolname compare to the state of the art?

\item \textbf{RQ2:} How does \toolname perform when there are multiple changes in the smart contract source code?

\item \textbf{RQ3:} How does the type of syntactic changes in file affect the performance of \textsc{SoliDiffy}? 

\item \textbf{RQ4:} How does \toolname perform against line differencing on commit history of real-world smart contracts? 
\end{itemize}

\subsection{Protocol for Research Questions}\label{sec:protocol}


\subsubsection{Protocol for RQ1}\label{sec:protocol-rq1}
To evaluate how well \toolname performs AST differencing for Solidity code, we compare it to an existing open-source tool, Difftastic~\cite{hughes_difftastic_2024}. Difftastic supports structural Solidity code differencing but does not generate fine-grained AST-based edit scripts. Difftastic generates a JSON edit script based on word-by-word replacement in the text. We compare it to the AST-based edit-scripts of \toolname.  
In our evaluation, we focus on the key metric of edit script length (lower is better) and successful completion rate of the differencing task (higher is better).

\subsubsection{Protocol for RQ2}
In this RQ, we analyze the effect of number of code differences on the performance of Solidity AST differencing tools, on the same dataset as RQ1.  

\subsubsection{Protocol for RQ3}

We generate mutations in existing Solidity contracts to evaluate the performance of \toolname and \othertool against specific types of changes.
The mutations used to generate the evaluation dataset range from simple syntax modifications to complex structural changes on Solidity smart contracts. We investigate the relationship between different operators and their edit distance.

\subsubsection{Protocol for RQ4}
We follow the same protocol as \emph{RQ1} with the difference of using a dataset of real-world commits in a popular smart contract project.

\subsection{Datasets}\label{sec:datasets}

\subsubsection{Seed Datasets}\label{sec:seed-datasets}

To follow the protocols of \emph{RQ1--3}, we need a dataset with a large number of Solidity source code files. For this, we seed synthetic modifications in  the DAppScan dataset~\cite{zheng_dappscan_2023, inpluslab_inpluslabdappscan_2024,morello_disl_2024} that contains a range of real-world audited smart contract projects. This dataset consists of \num{39904} Solidity source code files. From these \num{39904} files, we select \num{8102} files for the our experiment by removing all Solidity source files with duplicate names to ensure a diverse dataset and reduce the potential for redundancy that could undermine the validity of our evaluation.
The final dataset is available in a dedicated repository.\footnote{\href{https://github.com/SoliDiffy/SoliDiffyResults/tree/main/contracts/dataset}{https://github.com/SoliDiffy/SoliDiffyResults/tree/main/contracts/dataset}}

For \emph{RQ4}, we use the entire commit history of Uniswap v4 core smart contracts GitHub repository.\footnote{\href{https://github.com/Uniswap/v4-core}{https://github.com/Uniswap/v4-core}}

\subsubsection{Contract Pair Generation}\label{sec:contract-pair-generation}

To create contract pairs with varying levels of code alterations to the AST, one effective approach is to use a mutation testing tool~\cite{couplingEffectSWTesting}.  
Using mutations for difference generation provides automated, fine-grained code changes 
For this, we use mutation tool SuMo~\cite{barboni_sumo_2022, barboni_morenabarbonisumo-solidity-mutator_2024}, which is dedicated to Solidity. It contains 44 mutation operators that we all use\toolname\cite{phipathananunth_gambit_2022, noauthor_certoragambit_2024}.

We use a script\footnote{\url{https://github.com/mojtaba-eshghie/SoliDiffy/scripts/gen_diff_pairs.py}} to invoke SuMo from the command line, generating mutants for all files in the dataset using each available mutation operator. The process involves iterating through all \num{44} mutation operators, generating all possible mutations for each Solidity file, and creating up to \num{10} mutated versions per file. In some cases, the actual number of generated mutants is lower than 10 due to the limited mutation opportunities in some files. 
We note that some files in the dataset are incompatible with SuMo, causing crashes and preventing contract pair generation. 

In total, we generated \num{353262} contract pairs for differencing. The browsable version of these diff pairs is provided in our repository.\footnote{\url{https://github.com/SoliDiffy/SoliDiffyResults/tree/master/contracts/mutants}}\footnote{\url{https://solidiffy.github.io/}}

To generate contract pairs for differencing task of \emph{RQ4}, we processed the entire commit history of Uniswap v4 core project. The contract pair generation begins by retrieving the entire commit history of the \emph{main} branch in chronological order using \texttt{git log}, followed by identifying the specific Solidity files altered in each commit through \texttt{git diff-tree}. For each modified file, the script extracts the version of the file at both the current and previous state and stores them. Then, a \texttt{git diff} between these two versions is computed using \texttt{git diff}, and the differences are saved to a file. This dataset of contract pairs and their differencing results are available publicly at our results repository.\footnote{\url{https://github.com/SoliDiffy/SoliDiffyResults/tree/master/uniswap-v4-diffs}}

\subsection{Execution Environment}

We run the whole pipeline of the experiment on a system with an AMD EPYC 7742 64-core Processor and $528$ GB RAM. 
The total run time of the experiment was $6\hour 13\minute 43\second$. The differencing is parallelized based on the available number of CPU cores on the server.

\section{Experimental Results}\label{sec:results}

\subsection{Results for RQ1}\label{sec:res-rq1}

\begin{tcolorbox}[colback=blue!5!white, colframe=blue!75!black, fonttitle=\bfseries,
                  width=\linewidth, boxrule=0.2mm, rounded corners=southeast, boxsep=1mm, left=0.5mm, right=0.5mm, top=0.5mm, bottom=0.5mm]
\textbf{RQ1: } \rqOne
\end{tcolorbox}

\begin{table}[t]
\caption{Effectiveness of Solidity differencing tools on our large dataset (Section~\ref{sec:seed-datasets})}
\centering
\begin{tabular}{l c c}
\toprule
 & \textbf{\toolname} & \textbf{Difftastic} \\ 
\midrule
\multicolumn{1}{l}{{Total diffed pairs}} & \multicolumn{2}{c}{{\num{353262}}} \\ 
\midrule
Successfully diffed pairs & \num{339596} & \num{336331} \\ 
\bottomrule
\end{tabular}
\label{tab:effectiveness}
\end{table}

\begin{figure}[hbt!]
    \centering
    \includegraphics[width=1\columnwidth]{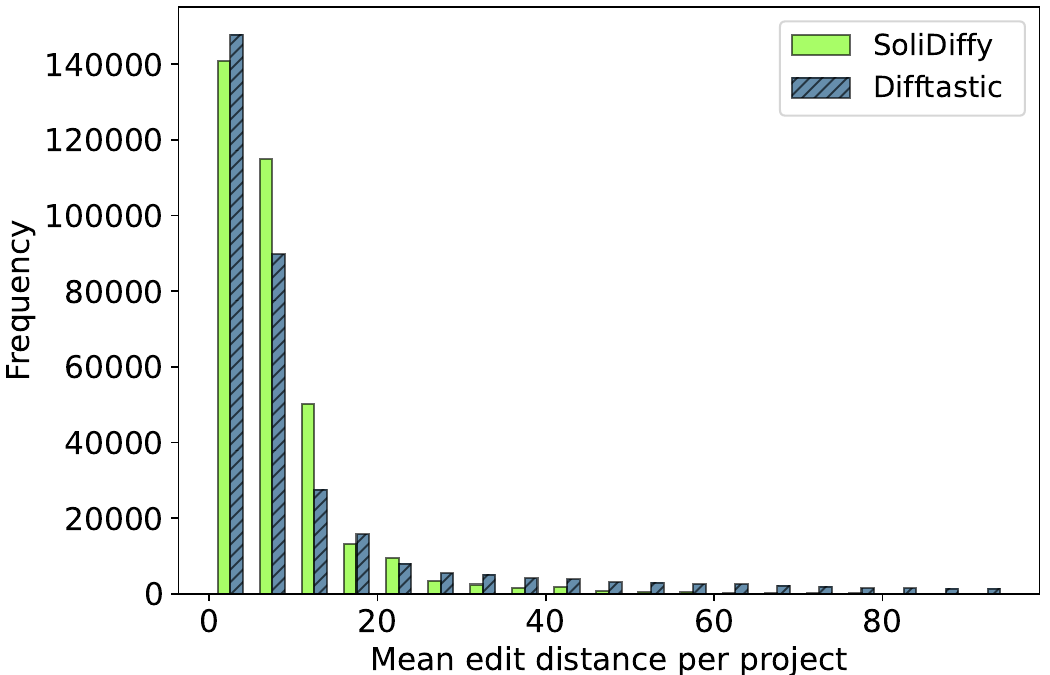}
    \caption{RQ1: Histogram of mean edit script lengths per project for \toolname and Difftastic across all diff pairs of the project ($n=\num{336331}$). The long tail of Difftastic's distribution is trimmed at \num{100} to fit the plot as it continues to more than \num{500}. Each pair of bars represents the frequency of projects falling within specific mean edit distance ranges.}
    \label{fig:histogram-per-project}
\end{figure}

We use \toolname and Difftastic to conduct a large-scale campaign of Solidity smart contract source code differencing (see Section~\ref{sec:experimental-protocols}). Figure~\ref{fig:histogram-per-project} shows the results of running the experiment, averaged across all diff pairs of each project (with varying contract pair modification severity and different types of modifications). We present the results as a side-by-side histogram. The green bars represent \textsc{SoliDiffy}, while the blue bars with a hatched pattern represent Difftastic. The y axis presents the frequency of edit script length that falls into a particular bin (x axis).  

The key result is that \textsc{SoliDiffy} produces shorter edit scripts, as witnessed by the bars for \toolname being consistently higher on the left side of the plot. Clearly, \toolname produces fewer edit actions across most contracts of the dataset.
In contrast, Difftastic's distribution is more spread out, with some edit scripts containining more than \num{80} changes and continuing to more than \num{500} which were trimmed to fit the plot. To ensure that the visual observations are statisically significant, we performed a Wilcoxon signed-rank test~\cite{wilcoxon1992individual} ($p < 0.001$) that shows difference between edit script length pairs over all projects is statistically significant.   

Moreover, as shown in Table~\ref{tab:effectiveness}, \toolname is able to successfully analyze more diff pairs than Diffstastic ($96.1\%$ vs. $95.1\%$). The main root cause of the crashes was syntax errors that were due to invalid syntax in mutated contracts.

\begin{tcolorbox}[colback=green!5!white, colframe=green!50!black, 
                  fonttitle=\bfseries,
                  width=\linewidth, boxrule=0.2mm, rounded corners=northwest, boxsep=1mm, left=0.5mm, right=0.5mm, top=0.5mm, bottom=0.5mm]
\textbf{Result for RQ1:} \textsc{SoliDiffy} outperforms Difftastic by producing shorter  edit scripts for Solidity smart contracts.  Additionally, \toolname successfully completed the analysis of a higher percentage of diffing tasks for contract pairs ($96.1\%$ vs.\ $95.1\%$).
\end{tcolorbox}

\subsection{Results for RQ2}\label{sec:res-rq2}
\begin{tcolorbox}[colback=blue!5!white, colframe=blue!75!black, 
                  fonttitle=\bfseries,
                  width=\linewidth, boxrule=0.2mm, rounded corners=southeast, boxsep=1mm, left=0.5mm, right=0.5mm, top=0.5mm, bottom=0.5mm]
\textbf{RQ2:} \rqTwo
\end{tcolorbox}

\begin{figure}[hbt!]
    \includegraphics[width=1\columnwidth]{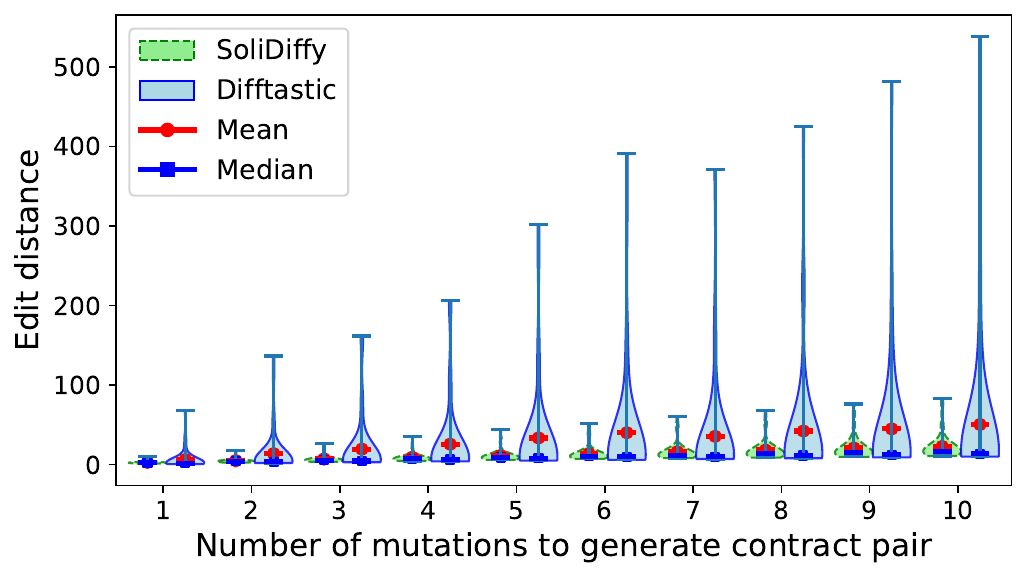}
    \caption{RQ2: Edit distances of \toolname and Difftastic per initial number of mutations. Triangles annotate the average edit distance.}
    \label{fig:rq2-result}
\end{figure}

\begin{figure*}[!t]
    \centering
    \includegraphics[width=1\textwidth]{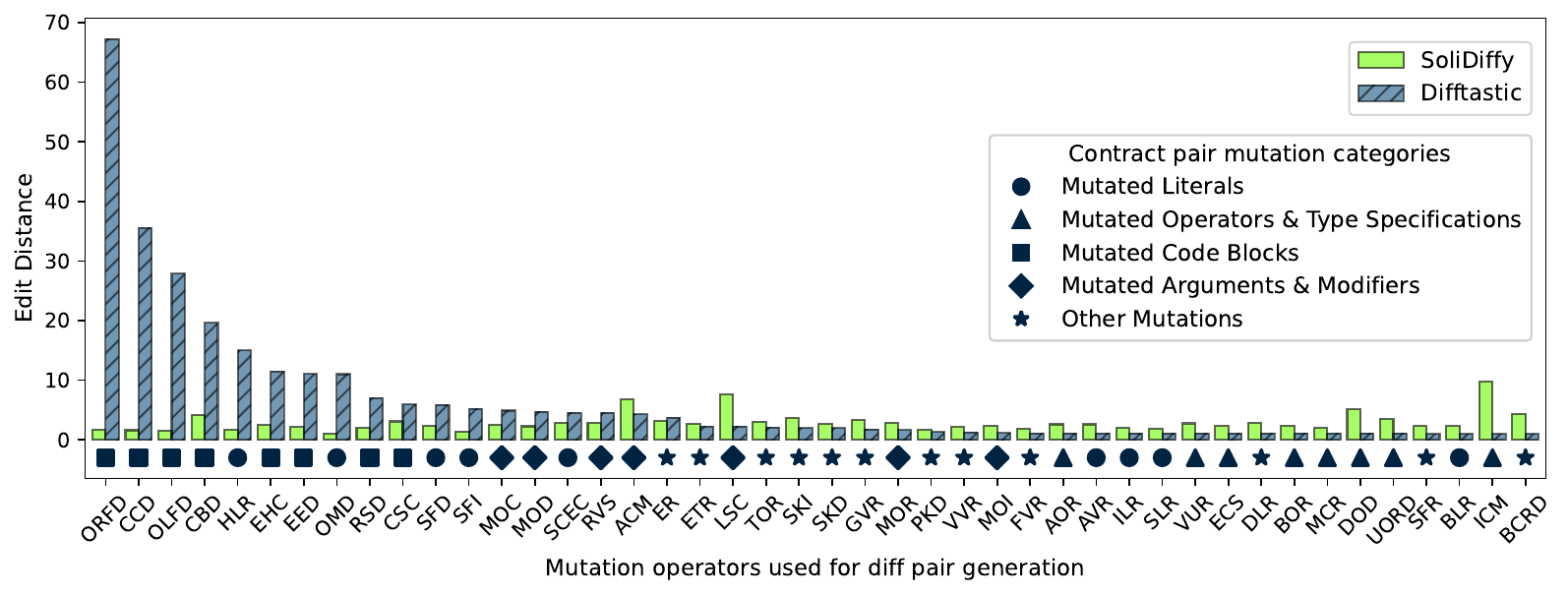}
    \caption{RQ3: Comparison of \toolname and \othertool edit distance split into different contract difference categories ($n = \num{36331}$). Different colors represent different numbers of mutations, ranging from $1$ to $10$. The bottom right legend describes the categories of modifications applied on each smart contract (Section~\ref{sec:datasets}).}
    \label{fig:edit-distance-by-categories}
\end{figure*}

Figure~\ref{fig:rq2-result} shows the results for the two differencing tools while running the diff pairs with the same operators repeatedly applied on the same Solidity contract (outlined in Section~\ref{sec:datasets}). 
The results are presented as a violin plot at which the width of the violin at different points shows the density of data. Key statistical markers, such as the mean and median in this plot highlight the skewness of the data. Peaks in the violin indicate where data clusters, and the tails represent outliers or extreme values. As Figure~\ref{fig:rq2-result} shows, \toolname is more dense towards the lower values of edit scripts, especially for lower number of mutations used for contract pair generation. For instance, differencing task performed on contract pairs which were the result of one and two mutations are very dense towards very low values of edit scripts for \textsc{SoliDiffy}. 

Furthermore, consistent extreme peaks in Difftastic violins especially when having more number of mutations, shows it tends to generate very long edit scripts for at least a consistent proportion of differenced contract pairs especially when the number of modifications are increase.  

While according to Figure~\ref{fig:rq2-result} \toolname exhibits lower values for the edit distance, we need a statistical test to confirm whether these differences are statistically significant.
Given that the distributions represented in the violin plots are not normally distributed, we employed the Kolmogorov-Smirnov (K-S) test~\cite{massey1951kolmogorov} to compare the two tools. The K-S test is suitable as it compares the entire cumulative distribution functions (CDFs) of two distributions, to detect differences not only in the central tendency but also in the overall shape, spread, and tails of the distributions. This is crucial because, as seen in the figure, Difftastic results show more variability and heavier tails compared to the more concentrated SoliDiffy distributions. The K-S test revealed statistically significant differences in all \num{10} mutation severity comparisons, with p-values consistently below the conventional threshold of 0.05 with the highest p-value being \num{0.005}. 
The results of the our K-S test align with the visual representation in Figure~\ref{fig:rq2-result}, where \toolname shows tighter distributions across all parameters, with the means consistently lower than Difftastic. \textsc{SoliDiffy}'s central tendency to lower edit distances, combined with the significantly different overall distribution shapes (as confirmed by the K-S test), provide compelling evidence of SoliDiffy's superior performance.

\begin{tcolorbox}[colback=green!5!white, colframe=green!50!black, 
                  fonttitle=\bfseries,
                  width=\linewidth, boxrule=0.2mm, rounded corners=northwest, boxsep=1mm, left=0.5mm, right=0.5mm, top=0.5mm, bottom=0.5mm]
\textbf{Result for RQ2:} \textsc{SoliDiffy} consistently produces smaller edit distances compared to Difftastic, regardless of the number of modifications in the smart contracts.
\end{tcolorbox}

\subsection{Results for RQ3}\label{sec:res-rq3}
\label{sec:res-per-mut}

\begin{tcolorbox}[colback=blue!5!white, colframe=blue!75!black, 
                  fonttitle=\bfseries,
                  width=\linewidth, boxrule=0.2mm, rounded corners=southeast, boxsep=1mm, left=0.5mm, right=0.5mm, top=0.5mm, bottom=0.5mm]
\textbf{RQ3:} \rqThree
\end{tcolorbox}

We analyze the results of applying diverse set of $44$ mutation operators on our dataset side-by-side for both \toolname and Difftastic. Figure~\ref{fig:edit-distance-by-categories} presents the performance difference between the two tools. 
The exact mutation operator and its category are written bellow each bar pair. 
The bars show the effect of mutation operator and its category used to create the smart contract pairs. Each bar represents the mean edit distance between the original smart contract and its respective modified version produced by applying the specific mutation operator only once. For instance, when diff pairs consist of mutated code blocks category, that is, when large blocks of code are added, moved, or removed from the code, the Difftastic edit distances are significantly larger than \textsc{SoliDiffy}'s results or any other type of modification.

As Figure~\ref{fig:edit-distance-by-categories} presents, the performance of \toolname and \othertool is considerably different in many cases. In most cases that \toolname performs better, it outperforms \othertool by a great margin. For instance, in all differencing tasks belonging to the mutated code blocks category where full blocks of code are manipulated, \toolname produces structurally meaningful edit scripts as opposed to \othertool which tends to produce edit scripts consisting of word-by-word additions or deletions. \toolname demonstrates a stronger performance where it matters most: in the cases where \othertool falls short, the discrepancies are notably more pronounced, highlighting the superior efficiency of \toolname in handling more complex differences, which are typical in real-world use-cases. 

For the cases where \toolname visibly produces longer edit scripts. For instance in the case of the textually minor mutation operator ICM (Increments Mirror) that changes an incrementing operator by swapping their two characters, \texttt{+=} becomes \texttt{=+}. The problem when representing this change in AST edit actions is that the ASTs generated from these two versions are very different. In the first (\texttt{+=}), an operator is applied to two values and the new value is written to one of them. In the other (\texttt{=+}), a value is simply set to a negative value. Difftastic's way of providing textual changes in this case provides a shorter edit distance as its edit script consists of adding \texttt{+} to an existing operand (\texttt{=}). Only the two characters that were swapped are displayed, and the resulting edit distance instead becomes the correct two per mutation. The same argument holds for all the instances that edit distances calculated by \toolname is higher than Difftastic.

For diff pairs belonging to argument/modifier and miscellaneous categories, \toolname and \othertool inconsistently outperform each other.
Our random sampling of diff pairs where \othertool producing smaller edit distance confirms that these belong to the cases where \othertool merely textually counts the number of add or removal of words in the Solidity contract.

We conducted a Wilcoxon signed-rank test~\cite{wilcoxon1992individual} to assess the statistical significance of the observed differences between \toolname and Difftastic across the \num{44} mutation operators. The test results show that for all \num{44} operators, the differences between \toolname and Difftastic were statistically significant
. This confirms that the visualized differences seen in Figure~\ref{fig:edit-distance-by-categories} are not due to random chance. 


\begin{tcolorbox}[colback=green!5!white, colframe=green!50!black, 
                  fonttitle=\bfseries,
                  width=\linewidth, boxrule=0.2mm, rounded corners=northwest, boxsep=1mm, left=0.5mm, right=0.5mm, top=0.5mm, bottom=0.5mm]
\textbf{Result for RQ3:} \toolname demonstrates superior performance in handling complex structural changes, particularly when large code blocks are modified. It can produce meaningful, concise edit scripts for diverse kinds of modification. 
\end{tcolorbox}

\subsection{Results for RQ4}\label{sec:res-srq4}

\begin{tcolorbox}[colback=blue!5!white, colframe=blue!75!black, 
                  fonttitle=\bfseries,
                  width=\linewidth, boxrule=0.2mm, rounded corners=southeast, boxsep=1mm, left=0.5mm, right=0.5mm, top=0.5mm, bottom=0.5mm]
\textbf{RQ4:} \rqFour
\end{tcolorbox}

Comparing Git line differencing which works on smart contract source level and not perform any structural (syntax-level) differencing, allows benchmarking \toolname against the most simplistic way of transforming one smart contract to another by merely removing lines and adding new lines. 
We define the edit distance of Git line differencing results as the total number of lines marked to remove and add.

\begin{figure}[hbt!]
    \includegraphics[width=1\columnwidth]{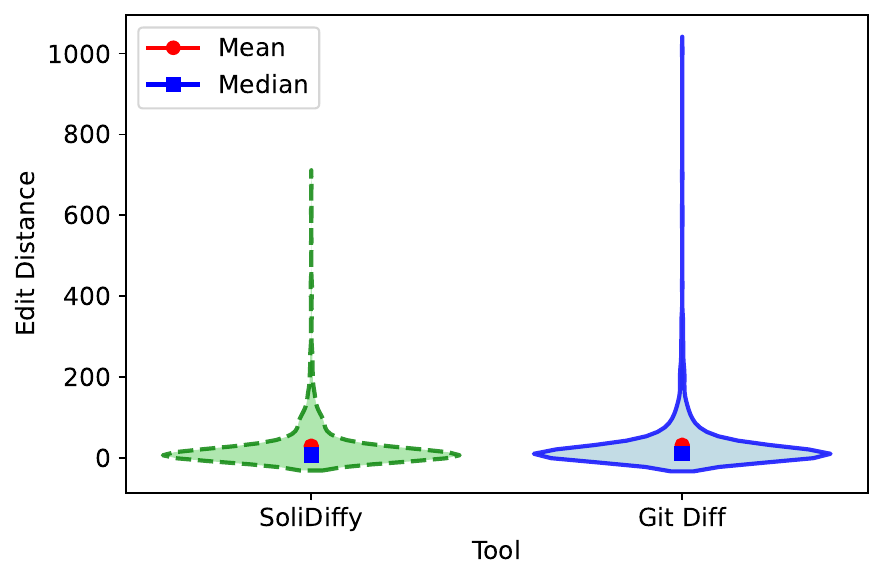}
    \caption{RQ4: Edit distances of \toolname and Difftastic for \num{925} smart contract pairs extracted from the commit history of Uniswap v4 core.}
    \label{fig:rq4-violin}
\end{figure}

Figure~\ref{fig:rq4-violin} shows the results of differencing task on \num{925} pairs of contract pairs with modifications from the commit history of Uniswap v4 core smart contracts. This figure visualizes \toolname and Git line differencing results using a violin plot for each tool.

While the violin shapes indicate that the two tools perform similarly on the lower end of the distribution, the spread of values—particularly the tail behavior—differs slightly between the tools. \toolname seems to consistently exhibit more compact results, while \texttt{git diff} demonstrates broader variability, especially in the higher ranges. 

To check the statistical significance of observation in Figure~\ref{fig:rq4-violin}, we conducted a Wilcoxon signed-rank test~\cite{wilcoxon1992individual} on edit script length of two tools generated on the same contract pairs (\num{925} pairs of edit distances). The result proved that distributions are indeed significantly different with a p-value of \num{0.01} rejecting the null hypothesis that there is no statistically significant difference between these two edit script length distributions.

\begin{tcolorbox}[colback=green!5!white, colframe=green!50!black, 
                  fonttitle=\bfseries,
                  width=\linewidth, boxrule=0.2mm, rounded corners=northwest, boxsep=1mm, left=0.5mm, right=0.5mm, top=0.5mm, bottom=0.5mm]
\textbf{Result for RQ4:} \textsc{SoliDiffy} provides structured and accurate differencing on real-world Solidity smart contract commits, significantly outperforming Git line differencing. The statistical tests confirm its reliability for practical use in tracking contract evolution.
\end{tcolorbox}

\section{Discussion}\label{sec:discussion}

\subsection{Lessons Learned}
Our experimental evaluation demonstrates that \textsc{SoliDiffy} offers significant improvements in edit script precision compared to Difftastic, particularly when handling complex code modifications, such as those involving large code blocks. By delivering shorter and more precise edit scripts, \textsc{SoliDiffy} provides developers with a clearer view of structural changes, reducing the cognitive load required for code reviews and audits. This is essential in the blockchain domain, where the immutability of deployed contracts necessitates rigorous pre-deployment analysis.

Our findings also revealed situations where \textsc{SoliDiffy} produced longer edit scripts than Difftastic. This was observed with certain low-impact syntax mutations where the distinction between AST nodes resulted in an increased edit distance. Such cases highlight that \textsc{SoliDiffy}'s more granular AST differencing may not always translate into shorter edit scripts, especially when the syntactic differences are minimal. Future work could explore hybrid approaches that incorporate both text-based and AST-based differencing to handle such cases more effectively.

The use of mutants as our primary dataset allowed for controlled evaluations, but it also introduced some limitations. Real-world smart contract updates often involve non-uniform changes that go beyond single syntactic mutations. The results of RQ4, using real-world data from Uniswap v4, suggest that \textsc{SoliDiffy} remains effective even in diverse commit histories, which indicates its robustness for practical applications. Further studies with varied real-world datasets could provide deeper insights into how well \textsc{SoliDiffy} performs in other scenarios, such as contract refactoring or collaborative development environments.

\begin{table*}[h]
\small
\centering
\caption{Summary of notable tools and research on Solidity source code differencing.}
\label{tab:related-tools}
\begin{tabular}{p{2.5cm} p{4cm} p{4cm} p{4cm}>{\centering\arraybackslash}p{1cm}}
\toprule
\textbf{Tool/Approach} & \textbf{Key Features} & \textbf{Differencing Technique} & \textbf{Limitations} & \textbf{Solidity} \\ \midrule
\textbf{srcML~\cite{maletic_supporting_2004}} & XML-based intermediate representation & Unix diff on XML & Limited evaluation and lacks structural analysis & \xmark \\ \midrule
\textbf{Dex~\cite{raghavan_dex_2004}} & Abstract Semantic Graphs (ASGs) instead of ASTs & Graph differencing & No move operation in edit scripts & \xmark \\ \midrule
\textbf{UMLDiff~\cite{xing_umldiff_2005}} & Reverse-engineered class models from Java code & Structural analysis with similarity scores & Limited to class structure, not applicable to all code changes & \xmark \\ \midrule
\textbf{Diff/TS~\cite{hashimoto_diffts_2008}} & Combines tree differencing with configurable heuristics & Tree differencing & No comparison to other similar tools in evaluation & \xmark \\ \midrule
\textbf{OperV~\cite{nguyen_operation-based_2010}} & Variable granularity using line and AST-based differencing & Line and AST differencing & Lacks comprehensive evaluation & \xmark \\ \midrule
\textbf{MTDIFF~\cite{dotzler_move-optimized_2016}} & Optimizes edit script length & Improved GumTree algorithm & Comparable failure rates with other tools & \xmark \\ \midrule
\textbf{IJM~\cite{frick_generating_2018}} & Merges nodes, prunes sub-trees for faster differencing & Improved GumTree algorithm & Lacks integration with mainstream Gumtree & \xmark \\ \midrule
\textbf{Matsumoto's Approach~\cite{matsumoto_beyond_2019}} & Splits AST nodes by line-based diff relevance & Line and AST-based differencing & Only focuses on improving specific troublesome actions & \xmark \\ \midrule
\textbf{Hunk-based AST Pruning~\cite{yang_pruning_hunks}} & Pruning ASTs based on unchanged lines & Line-based pruning for AST & Pruning has negligible impact on diff results  & \xmark \\ \midrule
\textbf{HyperAST~\cite{ledilvarec_hyperast}} & Single AST representing multiple file versions & AST storage across versions & Limited to AST construction optimization & \xmark \\ \midrule
\textbf{CLDiff~\cite{huang_cldiff_2018}} & Groups and links related edit actions & AST differencing with grouping & More coarse-grained, focused on grouping related changes & \xmark \\ \midrule
\textbf{SrcDiff~\cite{decker_srcdiff_2020}} & Heuristic-based matching, conversion rules & Heuristic-based differencing & Poor handling of complex updates in syntactic differencing & \xmark \\ \midrule
\textbf{Difftastic~\cite{hughes_difftastic_2024}} & Supports Solidity & Text changes differencing & Lacks concrete evaluation; uses text-based diffs; and only supports two edit actions & \checkmark \\ \midrule
\textbf{\toolname} & Move operation in edit scripts, RTED algorithm, Solidity-specific differencing & Top-down and bottom-up AST traversal and mapping to generate edit script for a diff pair, supports four edit actions & Issues with visual representation of diffs, edit script length for very small changes & \checkmark \\ 
\bottomrule
\end{tabular}
\end{table*}

\subsection{Threats to Validity}

\paragraph*{Construct Validity.} We evaluated \textsc{SoliDiffy} primarily through edit script length, a widely used metric in AST differencing research~\cite{falleri_fine-grained_2014, matsumoto_beyond_2019}. While shorter edit scripts generally indicate more precise differencing, this metric does not capture all aspects of quality, such as human interpretability or the practical utility of the generated edit scripts for downstream tasks.

\paragraph*{Internal Validity.} Our large-scale evaluation was performed on a synthetic dataset derived from the DAppScan~\cite{zheng_dappscan_2023} project. The use of mutation-based transformations ensured systematic variation but may not fully capture the complexity of organic contract evolution. To mitigate this, we supplemented our evaluation with real-world commits of Uniswap, but further studies across additional Solidity projects would provide deeper insights.

\paragraph*{External Validity.} The effectiveness of \textsc{SoliDiffy} was demonstrated across a diverse range of contract pairs, including both synthetic and real-world data. However, the generalizability of our results may be influenced by the dataset's composition. While Solidity is the dominant smart contract language, its ecosystem is rapidly evolving, and new language features or developer practices may introduce unforeseen challenges for AST differencing tools. Our open-source implementation allows the research community to extend our approach to newer Solidity versions and potentially adapt it for other smart contract platforms.

\paragraph*{Conclusion Validity.} The statistical tests performed in our study confirm the significance of \textsc{SoliDiffy}'s improvements over existing tools. However, statistical significance does not always translate directly to practical performance improvement. While \textsc{SoliDiffy} consistently produced shorter edit scripts and handled more contract pairs successfully, the impact of these improvements on downstream tasks, such as security auditing and automated program repair, requires further investigation.


\section{Related Works}\label{sec:related-works}

Here we provide a detailed review of the tools and research on source code differencing.

\subsection{Code Differencing From a Historical Perspective}


The srcML~\cite{maletic_supporting_2004} converts source code into an XML-based intermediate representation, retaining both the syntax and textual elements which allows for regenerating the original code. Unlike AST-based tools, it uses the Unix diff command on these XML files.
Dex introduced the use of Abstract Semantic Graphs (ASGs) instead of ASTs for C code, linking related nodes like variable references and declarations~\cite{raghavan_dex_2004}. Its differencing approach used graph rather than tree differencing, excluding the move operation from its edit scripts, but achieving a 95\% accuracy in detecting correct edit actions.
UMLDiff~\cite{xing_umldiff_2005} and Diff/TS~\cite{hashimoto_diffts_2008} focused on structural analysis of code changes. UMLDiff used class models reverse-engineered from Java source code to build change trees and calculated similarity scores to detect changes in the overarching class structure~\cite{xing_umldiff_2005}. Diff/TS, on the other hand, combined tree differencing with configurable heuristics to improve runtime and accuracy, incorporating all standard edit actions in its scripts~\cite{hashimoto_diffts_2008}.
OperV sought to offer variable granularity in version control systems, combining line-based differencing with AST-based matching, though its evaluation was limited compared to more modern approaches~\cite{nguyen_operation-based_2010}.

The aforementioned tools do not provide the neither fine-grained differencing capabilities required for smart contract languages such as Solidity.

Recent AST differencing tools, including GumTree~\cite{falleri_fine-grained_2014}, MTDIFF~\cite{dotzler_move-optimized_2016}, IJM~\cite{frick_generating_2018}, and the approach by Matsumoto et al.~\cite{matsumoto_beyond_2019}, focus on refining the generation of edit scripts (Section~\ref{sec:edit-scripts}).
One common method for evaluating edit script quality is by measuring its length~\cite{falleri_fine-grained_2014, matsumoto_beyond_2019, dotzler_move-optimized_2016}. Shorter edit scripts are generally preferred because they tend to contain fewer redundant operations and more closely align with the actual code modifications. Another approach is to count the number of matched nodes in the initial differencing step, which are not included in the edit script, providing insight into the tool’s effectiveness in detecting unchanged code structures. However, there are criticisms of these methods. For instance, focusing solely on reducing script length can sometimes lead to suboptimal results, as seen in tools like SrcDiff~\cite{decker_srcdiff_2020}. 

In addition to quantitative measures, qualitative evaluations through expert analysis are also commonly used. These smaller-scale assessments, as applied in studies of tools like GumTree~\cite{falleri_fine-grained_2014}, Matsumoto’s approach~\cite{matsumoto_beyond_2019}, and the differential testing conducted by Fan et al.~\cite{fan_differential_2021}, provide insights into the real-world usefulness of AST differencing tools.

\subsection{Gumtree Family of AST Differencing.}\label{cd-alg}
The Changedistiller algorithm~\cite{fluri_change_2007} is a foundational work in AST differencing, introducing a method to match identical nodes between two ASTs and generate an edit script. It built upon Chawathe's 1996 algorithm~\cite{Chawathe_change_detection}, optimizing it for source code by reducing edit script length by 45\%. This approach influenced many subsequent tools, including GumTree~\cite{falleri_fine-grained_2014}.
GumTree is particularly notable for its introduction of the \texttt{move} operation in edit scripts, which improves accuracy by grouping related changes. It uses a combination of top-down and bottom-up AST traversal and incorporates the RTED algorithm~\cite{pawlik_rted_2011} for generating mappings in smaller sub-trees. 
GumTree also supports hyperparameter tuning to optimize edit script length, as demonstrated by Martinez's Diff Auto Tuning (DAT) technique, which reduced script length
~\cite{martinez_hyperparameter_2022}. Additionally, GumTree can process general-purpose Tree-sitter CSTs by converting them into a format suitable for AST differencing~\cite{gumtreedifftree-sitter-parser_2023}.
MTDIFF~\cite{dotzler_move-optimized_2016} and Iterative Java Matcher (IJM)~\cite{frick_generating_2018} introduced improvements on built on top of GumTree algorithm but lack maturity and integration into the mainstream differencing tool.
While the aforementioned tools offer improvements for general-purpose languages, they lack specific adaptations for Solidity. \textsc{SoliDiffy} builds on this line of work by building on top of Gumtree's algorithms and ecosystem for Solidity smart contracts.

\subsection{Solidity Code Differencing}\label{sec:soldiff}

Research on code differencing specific to Solidity is limited. While line-based differencing can be used across languages, it lacks the precision needed for Solidity's unique syntax.
The only dedicated Solidity differencing tool in the literature is part of the Solidity Instrumentation Framework (SIF)~\cite{peng_sif_2019}. SIF uses AST-differencing, but its implementation is poorly documented and relies on an outdated AST format no longer supported by the Solidity compiler, making it unusable for newer code.
Outside academic literature, 
Difftastic~\cite{hughes_difftastic_2024} is the only other usable differencing tool that supports Solidity. It uses a Tree-sitter~\cite{TreesitterIntroduction} parser to generate side-by-side diffs or JSON output. However, it does not use traditional edit script generation, instead it focuses on concrete text changes (word-by-word), and as of now, no thorough evaluation of \othertool has been published..

\section{Conclusion}\label{sec:conclusion}

We introduced \textsc{SoliDiffy}, a novel AST differencing tool for Solidity smart contracts. \toolname provides fine-grained, accurate differencing, outperforming existing tools in both edit script quality and ability to handle complex syntactic changes. Our evaluation demonstrated that \textsc{SoliDiffy} supports AST differencing of  complex changes and real world contracts.
\toolname also gives an intuitive diff representation for developers. \textsc{SoliDiffy} sets a solid foundation for future enhancements in the field of smart contract analysis, such as incorporating semantic analysis or extending support to other blockchain languages.

\balance
\bibliographystyle{ieeetr}
\bibliography{refs}

\end{document}